# Characterization of Planar Cubic Alternative curve.

Azhar Ahmad[α], R.Gobithasan[γ], Jamaluddin Md.Ali[β],

[α]Dept. of Mathematics, Sultan Idris University of Education, 35900 Tanjung Malim, Perak, M'sia.
[γ]Dept of Mathematics, Universiti Malaysia Terengganu, 21030, Kuala Terengganu, M'sia.
[β]School of Mathematical Sciences, University Sains Malaysia, 11800 Minden, Penang, M'sia.

{azhar_ahmad@upsi.edu.my, gobithasan@umt.edu.my, jamaluma@cs.usm.my}

**Abstract**
In this paper, we analyze the planar cubic Alternative curve to determine the conditions for convex, loops, cusps and inflection points. Thus cubic curve is represented by linear combination of three control points and basis function that consist of two shape parameters. By using algebraic manipulation, we can determine the constraint of shape parameters and sufficient conditions are derived which ensure that the curve is a strictly convex, loops, cusps and inflection point. We conclude the result in a shape diagram of parameters. The simplicity of this form makes characterization more intuitive and efficient to compute.

## 1. Introduction

Fair curves are important in many computer aided design (CAD) and computer aided geometric design (CAGD) application. The reasons may aesthetic, so that the design of products may appear "visually pleasing" and in the "shape preserving" form. And it may be practical, where it is desirable that curves or surfaces can be generated. A generally accepted mathematical criterion for a curve to be fair is that it should have as few curvature extrema as possible, and it is often entail upon the convexity of the curve. Convex segment have be derived as a segment that having neither inflection point, cusp, or loop.

Characterization of the curve means to identifying whether a curve has any inflection points, cusps, or loops. Characterizing cubic curve has wide-ranging applications. For instance, in numerically controlled milling operations or in the design of highways, many of the algorithms rely on the fact that the trace of the curve or route is smooth; an assumption that is violated if a cusp is present. Inflection points often indicate unwanted oscillations in applications such as automobile body design and aerodynamics, and a surface that has a cross section curve possessing a loop cannot be manufactured.

Previous work in this area has been done by Su and Liu [4] have presented a specific geometric solution for the Bezier representation. By using canonical curve, Stone and DeRose [2] have characterized the parametric cubic curves. Walton and Meek [5] and followed by Habib and Sakai [3] have presented results on the number and location of curvature extrema upon cubic Bezier curve. All of them are using discriminant method in their solution. Yang and Wang (2004) have used the image of trochoids to investigate the occurrence of inflection and singularity of hybrid polynomial. For Bezier-like curve, Azhar and Jamaludin [1] have characterized rational cubic Alternative representation by using shoulder point methods and its only restricted for trimmed shape parameters.

This paper deals with cubic Alternative curve, which is the linear combination of the control points and basis functions that consist of two shape parameters. We apply the discriminant methods along with reparametrization methods to make a geometric characterization on untrimmed shape parameters of basis function.

The remaining part of this paper is organized as follows. In Section 2, we give a brief introduction of the cubic Alternative curve and some of its properties. The correspondence method and the used of reparametrization is described in Section 3. Our main result is shown in section 4, 5 and 6. Sufficient condition of inflection and singularity for nondegenerate curve is presented in Section 4, as well as degenerate curve in Section 6. Shape diagram of inflection and singularity, and some examples of correspondence curves is shown in Section 5.

## 2. Preliminaries

Referring [1], rational alternative cubic curve which have been use is given as

$$Z(t) = F_0(t)P_0 + F_1(t)P_1 + F_2(t)P_2 + F_3(t)P_3; \quad 0 \leq t \leq 1 \qquad (1)$$

with basis functions

$$\begin{aligned}
F_0(t) &= (1-t)^2(1+(2-\alpha)t) & F_2(t) &= \beta t^2(1-t) \\
F_1(t) &= \alpha(1-t)^2 t & F_3(t) &= t^2(1+(2-\beta)(1-t))
\end{aligned} \qquad (2)$$

Control points are denoted by $P_0, P_1, P_2, P_3$. While $\alpha$ and $\beta$ are shape parameters, in this study untrimmed shape parameters are used that was $\alpha, \beta \in$ . In general the cubic alternative curve possesses some interesting properties:

- Endpoint Interpolation Property - $Z(0) = P_0$ and $Z(1) = P_3$.
- Endpoint Tangent Property - $Z'(0) = \alpha(P_1 - P_0)$ and $Z'(1) = \beta(P_3 - P_2)$.
- Invariance under Affine Transformations.

This cubic alternative curve violated the Convex Hull property but it proved that for $0 \leq \alpha, \beta \leq 3$ satisfies thus property. The signed curvature $\kappa(t)$ of a plane curve $Z(t)$ is given by

$$\kappa(t) = \frac{Z'(t) \times Z''(t)}{\|Z'(t)\|^3}$$

Where $Z'(t)$ and $Z''(t)$ are first and second derivation of $R(t)$. Notation "×" is referring to outer product of two plane vectors. The signed radius at $t$ is the reciprocal of $\kappa(t)$. Its known that $\kappa(t)$ is positive signed when the curve segment bends to left at $t$ and its negative signed if it bends to right at $t$.

## 3. Description of Method

Now we consider a cubic Alternative curve as given by (1) and (2), it can be written as

$$Z(t) = (-1+t)^2(1+2t)P_0 + (3-2t)t^2 P_3 + (-1+t)^2 t\alpha(P_1 - P_0) + (-1+t)t^2 \beta(P_3 - P_2) \quad (3)$$

Assuming $(P_1 - P_0) = T_0$ and $(P_3 - P_2) = T_1$, so (3) can be rewritten as

$$Z(t) = (-1+t)^2(1+2t)P_0 + (3-2t)t^2 P_3 + (-1+t)^2 t\alpha T_0 + (-1+t)t^2 \beta T_1 \quad (4)$$

In this paper we use $P_1 = P_2 = H$, so that $\Delta Z = P_3 - P_0$ can be represented in terms of $T_0, T_1$;

$$\Delta Z = T_1 + T_0$$

So
$$P_3 = T_1 + T_0 + P_0 \quad (5)$$

And we assume that $T_0$ and $T_1$ are linearly independent, i.e., $T_0 \times T_1 (= \Gamma) \neq 0$. Substituting (5) into (4), give

$$Z(t) = P_0 + t\left(t(3-2\alpha) + t^2(-2+\alpha) + \alpha\right)T_0 + t^2\left(3 + t(-2+\beta) - \beta\right)T_1 \quad (6)$$

First and second derivative of (6) are

$$Z'(t) = (-1+t)\left(3t(-2+\alpha) - \alpha\right)T_0 + t\left(6 + 3t(-2+\beta) - 2\beta\right)T_1 \quad (7)$$
$$Z''(t) = 2\left(3 + 3t(-2+\alpha) - 2\alpha\right)T_0 + 2\left(3 + t(-2+\beta) - \beta\right)T_1 \quad (8)$$

A straight forward calculation gives

$$Z'(t) \times Z''(t) = -2\Gamma\left(t^2(3\alpha(-1+\beta) - 3\beta) - 3t\alpha(-2+\beta) - 3\alpha + \alpha\beta\right) \quad (9)$$

Let
$$\Phi(t) = \left(t^2(3\alpha(-1+\beta) - 3\beta) - 3t\alpha(-2+\beta) - 3\alpha + \alpha\beta\right) \quad (10)$$

The discriminant of quadratic function $\Phi(t)$ is given as

$$-3\alpha\beta(12 - 4(\alpha+\beta) + \alpha\beta) \quad (11)$$

For more intuitive and convenient analysis, we choose to apply reparametrization by using $t = \dfrac{u}{1+u}$, its mapped $t \in [0,1]$ onto $u \in [0,\infty)$. Hence, (9) can now be write as

$$Z'(u) \times Z''(u) = -\dfrac{2\Gamma}{(1+u)^2}\left(u^2\alpha(-3+\beta) - u\alpha\beta + (-3+\alpha)\beta\right) \tag{12}$$

Let

$$\Phi(u) = u^2\alpha(-3+\beta) - u\alpha\beta + (-3+\alpha)\beta \tag{13}$$

Consequently, the discriminant of $\Phi(u)$ is similar as shown by equation (11).

### 4. Characterization of the nondegenerate curve

Inflection point occur if and only if $Z'(u) \times Z''(u) = 0$. If the solutions are $r,s$ so

$$r,s = \dfrac{-\alpha\beta \pm \sqrt{-3\alpha\beta(12 - 4(\alpha+\beta) + \alpha\beta)}}{2\alpha(-3+\beta)} \tag{14}$$

Let $I = 12 - 4(\alpha+\beta) + \alpha\beta$, therefore (14) can be simplify as

$$r,s = \dfrac{\alpha\beta \pm \sqrt{-3\alpha\beta I}}{2\alpha(-3+\beta)} \tag{15}$$

Referring (14), it's clear that for planar cubic curve there are at most two inflection points, and the existence of inflection points and number of points are depend on $\alpha, \beta$ and $I$. First, we analyze the existence of real roots. For $r,s$ are complex roots then $\sqrt{-3\alpha\beta I}$ is not defined as $\alpha\beta I > 0$. This give us a result that, if $\alpha\beta > 0$ then $I > 0$, and if $\alpha\beta < 0$ then $I < 0$.

Second, for $r,s$ are real numbers, we probably have one or two inflection point. We make use of (13), quadratic equation as given below is use to determine the sign of $u$.

$$\left(u^2 - \dfrac{\beta}{(-3+\beta)}u + \dfrac{\beta(-3+\alpha)}{\alpha(-3+\beta)}\right) = 0 \quad \Leftrightarrow \quad u^2 + Au + B = 0 \tag{16}$$

Two negative roots imply no inflection point is obtained when A and B are both positive, this give $0 < \beta < 3$, $0 < \alpha < 3$. One positive root implies 1 inflection point is obtained when B is negative for A is positive. This give $0 < \beta < 3$, $\alpha < 0$, $I > 0$ and $0 < \beta < 3, \alpha > 3, I < 0$. One positive root implies 1 inflection point is obtained when both B and A are negative. This implies $0 < \alpha < 3$, $\beta < 0, I > 0$ and $0 < \alpha < 3, \beta > 3, I < 0$.

Finally, two positive roots imply 2 inflection points is obtained when A is negative and B is positive, give $\beta > 3, \alpha > 3, I < 0$, and $\beta > 3, \alpha < 0, I > 0$, and $\alpha > 3, \beta < 0, I > 0$.

The necessary condition for the occurrence of the inflection points along with numbers of inflection points of cubic Alternative curve can now be stated.

**Theorem 1.** For $Z(t)$ with untrimmed shape parameters, $\alpha, \beta \in$ , as defined by (3), the presence of inflection points on the cubic curve is characterized by the sign of $I = 12 - 4(\alpha + \beta) + \alpha\beta$

*Case 1* (2 inflection points): If $\beta > 3, \alpha > 3, I < 0$, and $\beta > 3, \alpha < 0, I > 0$, and $\alpha > 3, \beta < 0, I > 0$
*Case 2* (1 inflection point): If $0 < \beta < 3$, $\alpha < 0$, $I > 0$, and $0 < \beta < 3, \alpha > 3, I < 0$, and $0 < \alpha < 3$, $\beta < 0, I > 0$, and $0 < \alpha < 3, \beta > 3, I < 0$.
*Case 3* (none inflection point): If $\alpha\beta > 0, I > 0$, and $\alpha\beta < 0, I < 0$, and $0 < \alpha, \beta < 3$, $I < 0$.

Figure 1 showed the region for the numbers of inflection points $N_i$, $i = 0,1,2$ on restricted cubic Alternative curve, $0 \leq t \leq 1$. The shaded regions indicate two inflection points are in complex numbers (or none inflection points, $N_0$).

A necessary condition for existence of cusps is given by the vanishing of the first derivative vector $Z(t)$ in the given interval. Note that the quadratic polynomials $Z'(t) = (x'(t), y'(t))$ have the common zero(s). By using Sylvester's resultant of the above quadratic (7), we obtain

$$R(x'(t), y'(t)) = -3\Gamma^2 \alpha \beta I$$

From the definition of resultant, it is clear that the resultant will equal to zero if and only if $x'(t)$ and $y'(t)$ have at least one common root. Hence, a cusp occurs if $I = 0$. We obtain a common zero from substituting

$$\alpha = \frac{4(-3+\beta)}{(-4+\beta)} \quad (17)$$

into (7). The solution of thus quadratic gives

$$t = \frac{2(-3+\beta)}{3(-2+\beta)} \quad (18)$$

where cusp occur at that value. By reparametrization $t$ onto $u$, using $t = \frac{1}{1+u}$ we get

$$u = \frac{\beta}{2(-3+\beta)}, \tag{19}$$

Simple analyses of (18) upon value of $\beta$ give us following results;
- $\beta < 0 \Rightarrow u$ is positive so there exist a cusp
- $0 < \beta < 3 \Rightarrow u$ is negative, so no cusp
- $3 < \beta < 4 \Rightarrow u$ is positive so there exist a cusp
- $\beta > 4 \Rightarrow u$ is positive so there exist a cusp

*Remark;* Although it said that a cusp occurs if $I = 0$ but it not for $0 < \alpha, \beta \leq 3$, and it clear that $Z(t)$ is really quadratic curve for $(\alpha, \beta) = (2, 2)$.

By using $Z(u)$, $u \geq 0$ as defined from (6). Note that a loop occurs if and only if exist two roots of quadratic polynomials $p$ and $q$, where $\frac{Z(p) - Z(q)}{p - q} = 0$ and $p \neq q$. It gives a homogeneous system of equations in $MT_0 + NT_1 = 0$, where

$$M = -3p - p^2 - 3q - 10pq - q^2 - 3pq^2 + (p + q + 3pq - p^2q^2)\alpha \tag{20}$$

$$N = -3p - p^2 - 3q - 10pq - q^2 - 3pq^2 + (-1 + p^2q + pq(3+q))\beta \tag{21}$$

Since the matrix is nonsingular, we obtain $M = N = 0$, roots are

$$p, q = \frac{\alpha(8 + \alpha(-3+\beta) - 3\beta)\beta \pm (\alpha + \beta - \alpha\beta)\sqrt{\alpha\beta I}}{2\alpha(\alpha + (-3+\beta)\beta)} \tag{22}$$

And they defined when $\alpha\beta > 0, I > 0$ and $\alpha\beta < 0, I < 0$. We need to find the constrain of loop by finding the relation between shape parameter if one of the intercept point happen to be at $u = 0$. From (22), we obtained

$$(\alpha - 3\beta + \beta^2)(\beta - 3\alpha + \alpha^2) = 0 \tag{23}$$

These $(\alpha - 3\beta + \beta^2) = 0$ and $(\beta - 3\alpha + \alpha^2) = 0$ are parabola and showed in Figure 2 as $L_1$, $L_2$ respectively. The following theorem can now be stated.

**Theorem 2.** For $Z(t)$, with untrimmed shape parameters, $\alpha, \beta \in$ as defined by (3). The presence of singularity of the cubic Alternative curve is characterized by the sign of $I = 12 - 4(\alpha + \beta) + \alpha\beta$

Case 1 (Cusp): If $I = 0$ and $\alpha, \beta \in -[0,3]$.

Case 2 (Loop): If $\alpha\beta > 0, I > 0$, and $\alpha < 0, \beta > 0, I < 0, (\alpha - 3\beta + \beta^2) < 0$, and $\alpha > 0, \beta < 0, I < 0, (\beta - 3\alpha + \alpha^2) < 0$.

Case 3 (Quadratic): If $I = 0$, $(\alpha, \beta) = (2,2)$ no singularity, no inflection point.

## 5. Shape diagram

Figure 2, is shape diagram of the values of shape parameters consist in Alternative basis. It's happen that our results (shape diagram) are similar to the shape diagram obtained by Yang and Wang [6] for C-Bezier. The convex curve which is defined as not consists the inflection point, cusp and loop is show by region $C, H, V$ and $R$.

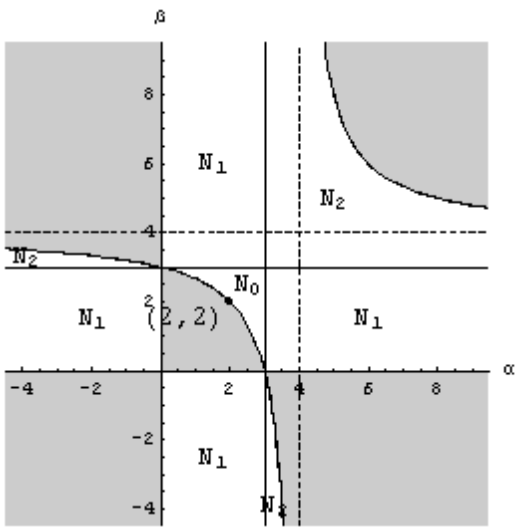 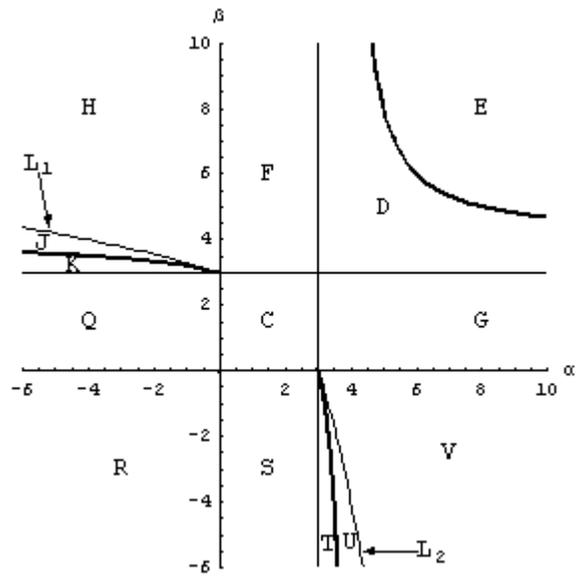

Figure 1: Region for numbers of inflection points.    Figure 2: Shape of inflection and singularity

Examples of these nine different shapes of cubic alternative curves are shown in Fig. 3. Details are as follow; (a) $(\alpha, \beta) \in C$, convex, (b) $(\alpha, \beta) \in D$, double inflections, (c) $(\alpha, \beta) \in I$ (upper branch), cusp, (d) $(\alpha, \beta) \in E$, loop, (e) $(\alpha, \beta) \in F$, one inflection point, (f) $(\alpha, \beta) \in V$, convex, (g) $(\alpha, \beta) \in U$, loop, (h) $(\alpha, \beta) \in S$, one inflection point, and (i) $(\alpha, \beta) \in R$, convex.

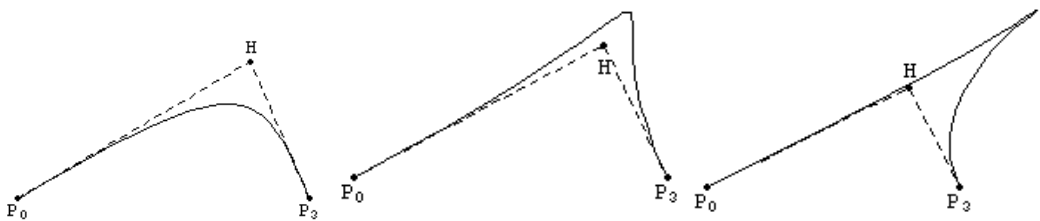

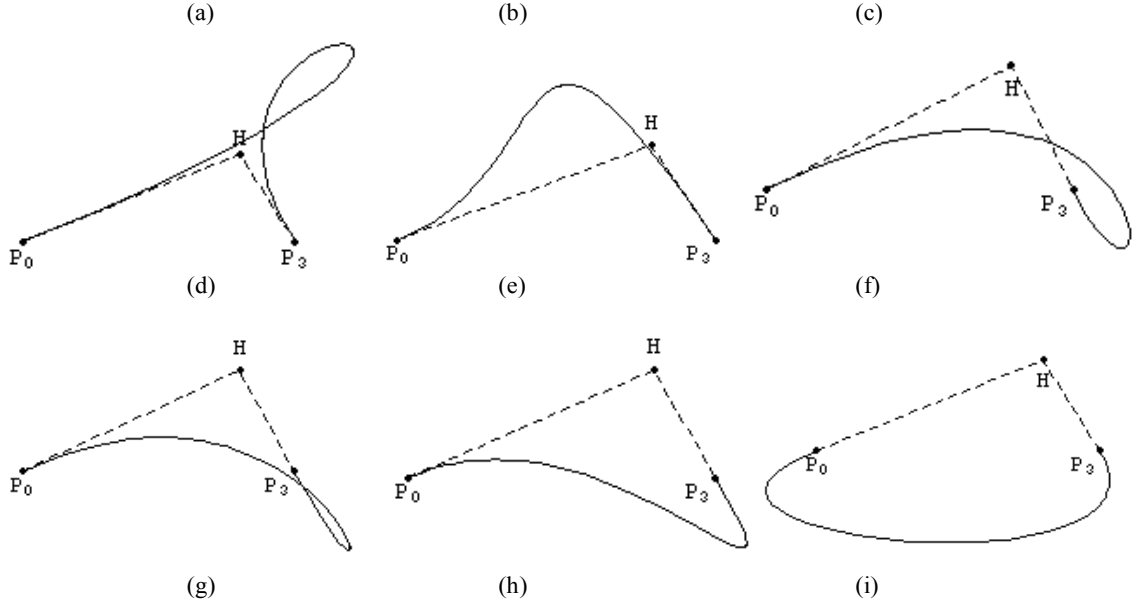

Figure 3: Examples of Cubic Alternative curve

## 6. Characterization of the degenerate curve

Finally, we will discuss on degenerate cubic which involving the parallel of the end tangent. We omitted the case where $P_0, H, P_3$ are in collinear position because the result is only the line segment. We consider cubic Alternative given by (3). Let assign

$$\begin{aligned} P_1 - P_0 &= \mu Z_0 \\ P_3 - P_2 &= -\nu Z_0 \\ P_2 - P_1 &= m Z_S \end{aligned} \qquad (24)$$

$P_1 - P_0$ is parallel to $P_3 - P_2$, $\mu, \nu, m$ are arbitraries and $Z_0$ are unit vector tangent of endpoint. Where $Z_S$ is unit vector along $P_2 - P_1$. $P_3$ in the terms of unit vectors is

$$P_3 = P_0 + (\mu\alpha - \nu\beta)Z_0 + mZ_S \qquad (25)$$

Let $a = \mu\alpha$, $b = \nu\beta$. Substituting (24),(25) into (3) yield

$$Z(t) = P_0 + m(3-2t)t^2 Z_S + t\left(a(1+t-t^2) + b(-2+t)t\right)Z_0 \qquad (26)$$

By reparametrization, $t$ onto $u$ therefore we rewrite (26) as

$$Z(u) = P_0 + \frac{m(1+3u)}{(1+u)^3} Z_S + \frac{(-b(1+2u) + a(1+3u+u^2))}{(1+u)^3} Z_0 \qquad (27)$$

Analogously as previous method on non degenerated curve, we obtained

$$Z'(u) \times Z''(u) = -\frac{6m(b+au^2)}{(1+u)^2} \Upsilon, \qquad (28)$$

where $\Upsilon = Z_0 \times Z_S$.

Occurrence of inflection points are if and only if $Z'(u) \times Z''(u) = 0$. So the solutions are

$$u = \pm \frac{i\sqrt{b}}{\sqrt{a}}. \qquad (29)$$

Observe that there is only a single inflection point exist for $ab < 0$.

No cusp is exist in this case since resultant of $Z'(t) = (x'(t), y'(t))$ are $R(x'(u), y'(u)) = -36abm^2 \Upsilon^2 \neq 0$

Again, loop occurs if and only if exist two roots of quadratic polynomials $p$ and $q$, where $\frac{Z(p) - Z(q)}{p - q} = 0$ and $p \neq q$. A homogeneous system of equations in $MT_0 + NT_1 = 0$ for this case is

$$\begin{aligned} M &= p^2\left(b + 2bq - a(1 + 3q + q^2)\right) + p\left(b(3 + 7q + 3q^2) - a(2 + 7q + 3q^2)\right) \\ &\quad - aq(2+q) + b(1 + 3q + q^2) \\ N &= -mp^2(1 + 3q) - mp(3 + 10q + 3q^2) - mq(3 + q) \end{aligned} \qquad (30)$$

Since $M = N = 0$ for occurrence of the loop, hence the roots of quadratic equations from (30) are

$$p, q = \frac{4ab \pm \sqrt{3}\sqrt{ab(a+b)^2}}{a(a - 3b)} \qquad (31)$$

It clear that $p, q$ are not defined if $ab < 0$. Now we consider that for $ab > 0$, a quadratic equation in terms of $u$ can be obtain from (31) as

$$u^2 - \frac{8b}{(a-3b)} u + \frac{b(-3a+b)}{a(a-3b)} = 0 \Leftrightarrow u^2 + Au + B = 0 \qquad (32)$$

Our attention is to prove that if a loop is exist so there are two positive roots of (32) can be obtain. First, we gain that the condition require for $A<0$ are $a,b>0$, $0<b<\frac{a}{3}$ and $a,b<0$, $\frac{a}{3}<b<0$. Now from $B$, analysis showed that if $a,b>0$, $A<0$ so for $B>0$, we yield $b>3a$. This result is contradict with $a,b>0$, $0<b<\frac{a}{3}$ for $A<0$. And if $a,b<0$, $A<0$ so for $B>0$, we yield $b<3a$, this result is also contradict with $\frac{a}{3}<b<0$ for $A<0$. As the conclusion, $B$ cannot be positive if $A<0$. So there are no two values of $u$ in positive sign, hence it implies that no loop for this case.

### 7. Conclusion

We investigated the existence of inflection points, cusps, and loops in cubic Alternative curve by using the discriminant and reparametrization methods, and finally computed a simple shape diagram. It's clear that the conditions of shape parameters consist in basis function are free from the position of control points. Finally, it would be interesting to find how these methods are useful for characterizing the shapes of curve in more general case.